\documentclass[sigconf]{acmart}
\usepackage{multirow}
\usepackage{tabulary}
\usepackage{subfigure}
\usepackage{graphicx}

\AtBeginDocument{%
  \providecommand\BibTeX{{%
    \normalfont B\kern-0.5em{\scshape i\kern-0.25em b}\kern-0.8em\TeX}}}

\copyrightyear{2023} 
\acmYear{2023} 
\setcopyright{acmlicensed}\acmConference[CIKM '23]{Proceedings of the 32nd
ACM International Conference on Information and Knowledge
Management}{October 21--25, 2023}{Birmingham, United Kingdom}
\acmBooktitle{Proceedings of the 32nd ACM International Conference on
Information and Knowledge Management (CIKM '23), October 21--25, 2023,
Birmingham, United Kingdom}
\acmPrice{15.00}
\acmDOI{10.1145/3583780.3615002}
\acmISBN{979-8-4007-0124-5/23/10}
\begin{document}

\title[Improving Customer Lifetime Value Modelling via Expert Routing and Game Whale Detection]{Out of the Box Thinking: Improving Customer Lifetime Value Modelling via Expert Routing and Game Whale Detection}


\author{Shijie Zhang}
\affiliation{%
  \institution{Tencent}
  \city{Shenzhen}
  \country{China}}
\email{julysjzhang@tencent.com}
\authornote{Corresponding authors}

\author{Xin Yan}
\affiliation{%
  \institution{Tencent}
  \city{Shenzhen}
  \country{China}}
\email{cyndiyan@tencent.com}

\author{Xuejiao Yang}
\affiliation{%
  \institution{Tencent}
  \city{Shenzhen}
  \country{China}}
\email{jerrieyang@tencent.com}

\author{Binfeng Jia}
\affiliation{%
  \institution{Tencent}
  \city{Shenzhen}
  \country{China}}
\email{vincejia@tencent.com}

\author{Shuangyang Wang}
\affiliation{%
  \institution{Tencent}
  \city{Shenzhen}
  \country{China}}
\email{feymanwang@tencent.com}








\begin{abstract}

Customer lifetime value (LTV) prediction is essential for mobile game publishers trying to optimize the advertising investment for each user acquisition based on the estimated worth. In mobile games, deploying microtransactions is a simple yet effective monetization strategy, which attracts a tiny group of game whales who splurge on  in-game purchases. The presence of such game whales may impede the practicality of existing LTV prediction models, since game whales' purchase behaviours always exhibit varied distribution from general users. Consequently, identifying game whales can open up new opportunities to improve the accuracy of LTV prediction models. However, little attention has been paid to applying game whale detection in LTV prediction, and existing works are mainly specialized for the long-term LTV prediction with the assumption that the high-quality user features are available, which is not applicable in the UA stage. In this paper, we propose ExpLTV, a novel multi-task framework to perform LTV prediction and game whale detection in a unified way. In ExpLTV, we first innovatively design a deep neural network-based game whale detector that can not only infer the intrinsic order in accordance with monetary value, but also precisely identify high spenders (i.e., game whales) and low spenders. Then, by treating the game whale detector as a gating network to decide the different mixture patterns of LTV experts assembling, we can thoroughly leverage the shared information and scenario-specific information (i.e., game whales modelling and low spenders modelling). Finally, instead of separately designing a purchase rate estimator for two tasks, we design a shared estimator that can preserve the inner task relationships. The superiority of ExpLTV in terms of its LTV prediction and game whale detection effectiveness is further validated via extensive experiments on three industrial datasets.
\end{abstract}




\maketitle

\section{Introduction}
\begin{figure}
    \centering
    \includegraphics[scale=0.34]{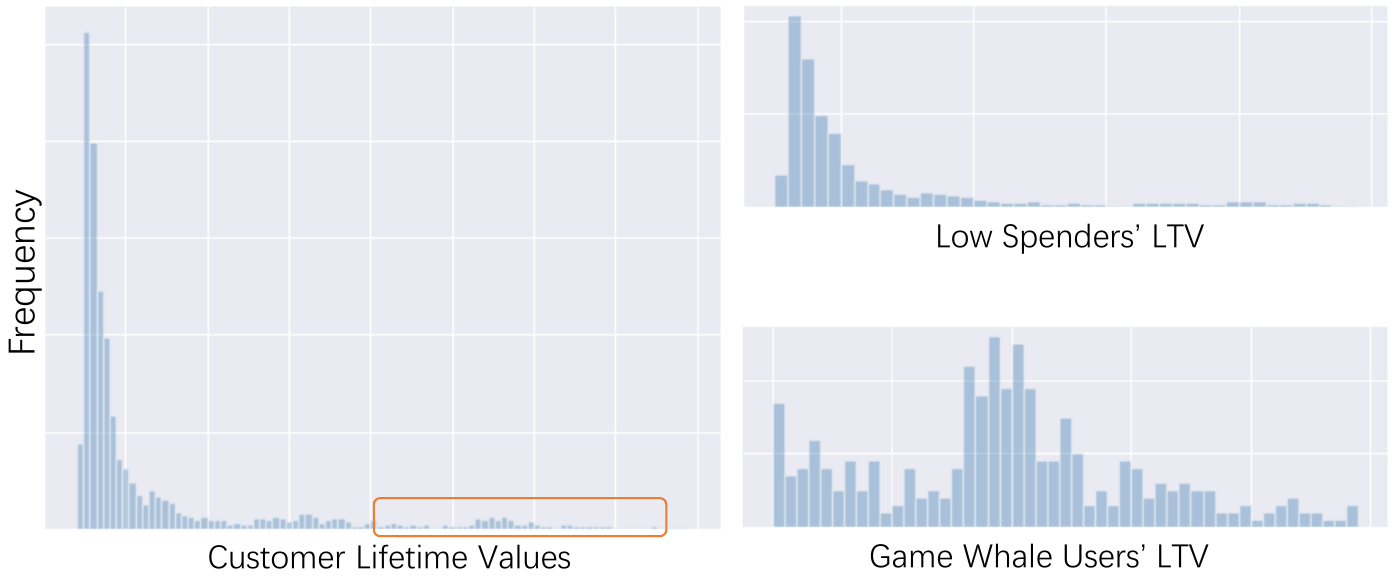}
    \vspace{-0.5em}
    \caption{The distribution of LTV values on GAME A.}
    \vspace{-1.5em}
    \label{fig:intro}
\end{figure}
For top-grossing games, a well-rounded user acquisition (UA)~\cite{su2022data} strategy is privilege to grow their user base. Without an effective strategy, the game publishers will miss out on swathes of opportunities to convert more valuable users. The easiest way to get the attention of new users on online services for promoting games is paid advertising. Most of social media (e.g., Sina Weibo) with massive active users own real-time bidding services for online advertising to deliver paid ads to their users, and thus it is an effective way to connect with target consumers and promoting games. 

Customer Lifetime Value (LTV)~\cite{hoekstra1999lifetime,berger1998customer,jain2002customer,segarra2022engaging} which refers to the total worth that a company can attribute to a customer over a specific period $T$. It is not only essential to estimate the long-term revenue from an established customer for personalized customer relationship management (CRM)~\cite{anshari2019customer, kumar2018customer}, but also allowing to predict the short-term worth of a new customer for adjusting the budget of advertising investment. For each ad impression, the advertisers can compute an early estimation of the customer's LTV value in the target games, and then the predicted values will be used to adjust the final bidding price, so as to enhance the success of bidding valuable users and optimize the marketing budget. In this context, LTV prediction accuracy means that the predicted LTV values should be sorted in a manner consistent with the true LTV values, and inaccurate LTV predictions will lead to an extra market budget and decreased success of UA. In this regard, many sorts of LTV prediction models have been developed for various applications. The early LTV prediction methods assume the purchase behaviors as a probability distribution, and then employ the probabilistic generative models for predicting the future purchase values and customer churn~\cite{gupta2006modeling,fader2005rfm,fader2005counting,schmittlein1987counting}. Another line of research proposes machine learning-based models to learn a mapping between hand-crafted features and monetary value of game players ~\cite{drachen2018or, vanderveld2016engagement, gupta2006modeling}. For example,~\cite{vanderveld2016engagement} is designed to predict future value on an individual user basis with a random forest model in Groupon. Recently, the prominent development of deep learning-based techniques brings more opportunities to improve the performance of LTV prediction models.~\cite{chen2018customer} adopts a convolutional neural network to automatically learn temporal representations for LTV prediction. To handle the volatile and sparse data of monetary value,~\cite{xing2021learning} adopts Temporal Trend Encoder and Graph Attention Network~\cite{velivckovic2017graph} to learn the temporal and structure representation respectively.~\cite{li2022billion} proposes an industrial solution in KUAISHOU, which predicts users' DAU via multiple distribution models to deal with the complex distribution.

Despite the efficacy of existing LTV predictions in many real-world applications, the majority of these methods are inapplicable in our scenarios. Most of these methods are only focused on the long-term LTV predictions with the assumption that the rich and high-quality user features are available. For example, in~\cite{xing2021learning}, to outright predict day $30$ LTV, the proposed model requires access to a large number of users with day $120$ LTVs and user features available. Due to app updates, user interest changes and time sensitivity, it is of great practical significance to construct a short-term LTV prediction model for game promotion. Moreover, as the gaming industry has progressed, microtransactions have been a regular fixture in gaming to boost revenues and improve user life cycle, especially for those free-to-play games. Gamers can purchase additional in-game items to decorate their character, upgrade weapons, or gain extra perks. Provided such microtransactions tends to attract many Game Whales (i.e., high spenders) who splurge on in-game purchases\cite{dreier2017free, britt2021waifus}. In the presence of such unpredictable and extreme purchase behaviours, these aforementioned LTV prediction models are subject to different levels of performance drop in the mobile game context, since the models are designed without the awareness of game whales and are sensitive to large values. 

In mobile games, the game whales (GW) represent the smallest percentage of users who are responsible for up to $50\% - 80\%$ in revenue sales. Thus, it is imperative to spot these game whales in the UA stage. Winning these customers can create a positive feedback loop, where they bring in more profit for game operations. Despite the importance of detecting game whales, most existing solutions~\cite{chen2018customer} straightforwardly use predicted LTV values as the dominant indicators to detect those high-value users. Since the game whales are rare and corresponding LTV values are largely deviated from general users, the pure LTV prediction models trained by heavily imbalanced dataset cannot perform well on the extremely minority labels, and thus lead to inferior performance in game whale detection task. To provide a proof-of-concept, in Figure \ref{fig:intro}, we show the LTV distribution of GAME A. Clearly, the game whales' LTV values demonstrate a different long-tailed distribution from that of low spenders. Moreover, directly using conventional deep-learning based models in game whale detection task will suffer from \textbf{Sample Selection Bias (SSB)}~\cite{zadrozny2004learning} and \textbf{Data Sparsity (DS)}!\cite{chen2022thinking} problems. SSB problem exists due to a flaw in the training process where a subset of the data is under-sampled, and thus significantly bias the estimates in the inference space. Specifically, GW detection models are trained on dataset composed of paying users, while are utilized to make inference on all samples of convert users. Such biased training will cause severe performance drop in online services. In addition, the game whales with outstanding LTV values are far less than general users, and thus a new paradigm is desired to effectively counteract the long-standing problem of DS problem.  

To this end, we aim to propose a novel multi-task framework named ExpLTV, which makes full use of GW detection to boost LTV prediction accuracy. Instead of directly using simple constraints to divide users into game whales and general users, we innovatively design a mapping function to calculate the user's probability of being game whales and transfer the binary classification task into a regression task. In this way, the specifically designed detector can not only capture the intrinsic order in accordance with the monetary value, but also generate the probability of being low spenders and high spenders to bucket users for customized game marketing. In addition, to eliminate the aforementioned SSB and DS problems simultaneously, we form a new sequential behavior "$convert\rightarrow purchase\rightarrow game whales$" where the pre-actions are more abundant, and then propose two auxiliary tasks of predicting the purchase-through rate (PTR) and purchase-through\&game whale rate (GWPTR) trained via multi-task learning. As such, our model can leverage extra supervisory signals from auxiliary tasks and all samples over the entire space. To allow for accurate LTV prediction, we put forward two novel LTV experts to model users' monetary values instead of one LTV model to serve all users. Inspired by ~\cite{wang2019deep}, in each LTV expert, we model the distribution of LTV as the zero-inflated lognormal (Ziln) distribution, which consists of three elements, namely purchase rate, mean, and standard deviation parameters. Correspondingly, our LTV prediction model is optimized by zero-inflated lognormal loss, which can minimize the model's sensitivity of the large values from game whales. Moreover, we take GW detector as the gating network to route users into the right expert. By this way, each expert focuses on a specific type of users thus accommodating different scenarios (i.e., high spenders and low spenders modelling) and maintaining discriminative characteristics. Finally, the purchase rate estimator trained in the LTV component is also used in GW detection, which can capture task-relatedness. As a result, these two components are closely hinged and make their complementary merits. 

Overall, we summarize our contributions in the following: 
\begin{itemize}
    \item To the best of our knowledge, we are the first to introduce the idea of investigating the mutually beneficial relationship between the LTV prediction task and the game whale detection task. 
    \item We propose ExpLTV, a novel multi-task framework that bypasses the shortcomings of conventional LTV prediction methods, allowing the game whale detector as a gating network to allocate the users to be trained via the right pattern of LTV experts assembling. 
    \item To alleviate the SSB and DS problems in GW detection, we form a new sequential behavior "$convert\rightarrow purchase\rightarrow game whales$", and then transfer GW detection into two auxiliary tasks (i.e., PTR and GWPTR). Furthermore, the PTR estimator shares the same model parameters with the purchase rate estimator in the LTV model, which can capture the task relatedness, and thus boost the corresponding performance. 
    \item Extensive experiments are conducted on three industrial datasets to evaluate the performance of ExpLTV, and the experimental results show that ExpLTV achieves superior performance in both LTV prediction and GW detection tasks.
\end{itemize}

\section{Preliminary}
\label{df:definition}
In this section, we first revisit key concepts and then mathematically formulate our research problems. Notably, vectors and matrices are denoted by bold lowercase and bold uppercase letters respectively, and sets are calligraphic uppercase letters. 

\textbf{Definition 1} (Customer Lifetime Value): $LTV^T_u$ is the total worth to a mobile game of a customer over a specific period $T$. In this work, $T$ is a constant value and thus we let $LTV_u$ replace $LTV^T_u$ for simplicity.

\textbf{Definition 2} (Game Whale): In mobile games, the game whales represent a group of users who are responsible for the majority of in-app's revenue. Formally, we use $g_u$ to denote the identity label of user $u$. In our case, if user $u$'s $LTV \geq R$, then user $u$ is defined as a game whale with $g_u = 1$. As the purchase behavior of the game whale should be observed, we let $s_u = 1$ denote it.  

\textbf{Definition 3} (Game Whale and Purchase Probability): For user $u$, let $p_u^{gwptr}$ denote user $u$'s probability of purchase and being a game whale. Specifically, we design a mapping function which calculates $p_u^{gwptr}$ based on the available LTV values:
\begin{equation}
    p_u^{gwptr} = p(g_u = 1, s_u = 1|\mathbf{x}_u) = 1-e^\frac{-LTV_u}{R} 
\end{equation}
Note that $p_u^{gwptr} = 0$, if and only if $LTV_u = 0$.

\textbf{Definition 4} (Conditional Game Whale Probability): The aim of GW detection is estimating $p_u^{gw}$, which denotes the conditional probability of being detected as a game whale, given that $u$'s purchase behavior is observed. Given a convert user $u$, the $p_u^{gw}$ can be represented as:
\begin{equation}
     p_u^{gw} = p(g_u=1| s_u=1,\mathbf{x}_u)
\end{equation}
Note that $p_u^{gw}$ as the conditional probability which is usually obtained via Bayes' theorem. 

\textit{Task 1}. \textbf{Game Whale Detection}: For each user $u \in \mathcal{U}$, we construct a feature vector $\mathbf{x}_u \in \mathbb{R}^{m}$ where $\mathbf{x}_u$ consists of dense features (e.g., purchase frequency), categorical features (e.g., gender) and sequence features (e.g., purchase behaviours). Given a set of samples $\mathcal{D} = \{(\mathbf{x}_u, p^{gwptr}_u) \in \mathbf{X} \times [0, 1]: u \in \mathcal{U}\}$, game whale detector is trained to estimate users' probability $\hat{p}^{gw}_u \in [0, 1]$ in terms of the definition. With the sortable probability computed, the GW detection task aims to recommend a set of game whales by selecting $K$ top-ranked users w.r.t. $\hat{p}_u^{gw}$: It can be formulated as:
\begin{equation}
    GWD(\mathcal{D}, \Theta_1) = \{u_i| \hat{p}^{gw}_{u_i} \text{ is top-K in } \{\hat{p}_{u_i}^{gw}\}_{u_i \in \mathcal{U}}\}
\end{equation}
where $\Theta_1$ represents the parameters of the GW Detector. Note that Section 3.3 will introduce the reasons why we use $p_u^{gwptr}$ as the label in this task rather than $g_u$. 

\textit{Task 2}. \textbf{Customer Lifetime Value (LTV) Prediction}: Given a set of samples $\mathcal{D} = \{(\mathbf{x}_u, LTV_u) \in \mathbf{X} \times N^+\cup\{0\}: u \in \mathcal{U}\}$, we aim to predict $LTV_u$ for user $u$ who is a new register or new returning user, which can be represented as:
\begin{equation}
    \widehat{LTV}_u = f(\mathbf{x}_u|\mathcal{D}, \Theta_2),
\end{equation}
where $\Theta_2$ denotes the parameters of the LTV predictor.

\section{Methodology}
\begin{figure*}
    \centering
    \includegraphics[scale=0.52]{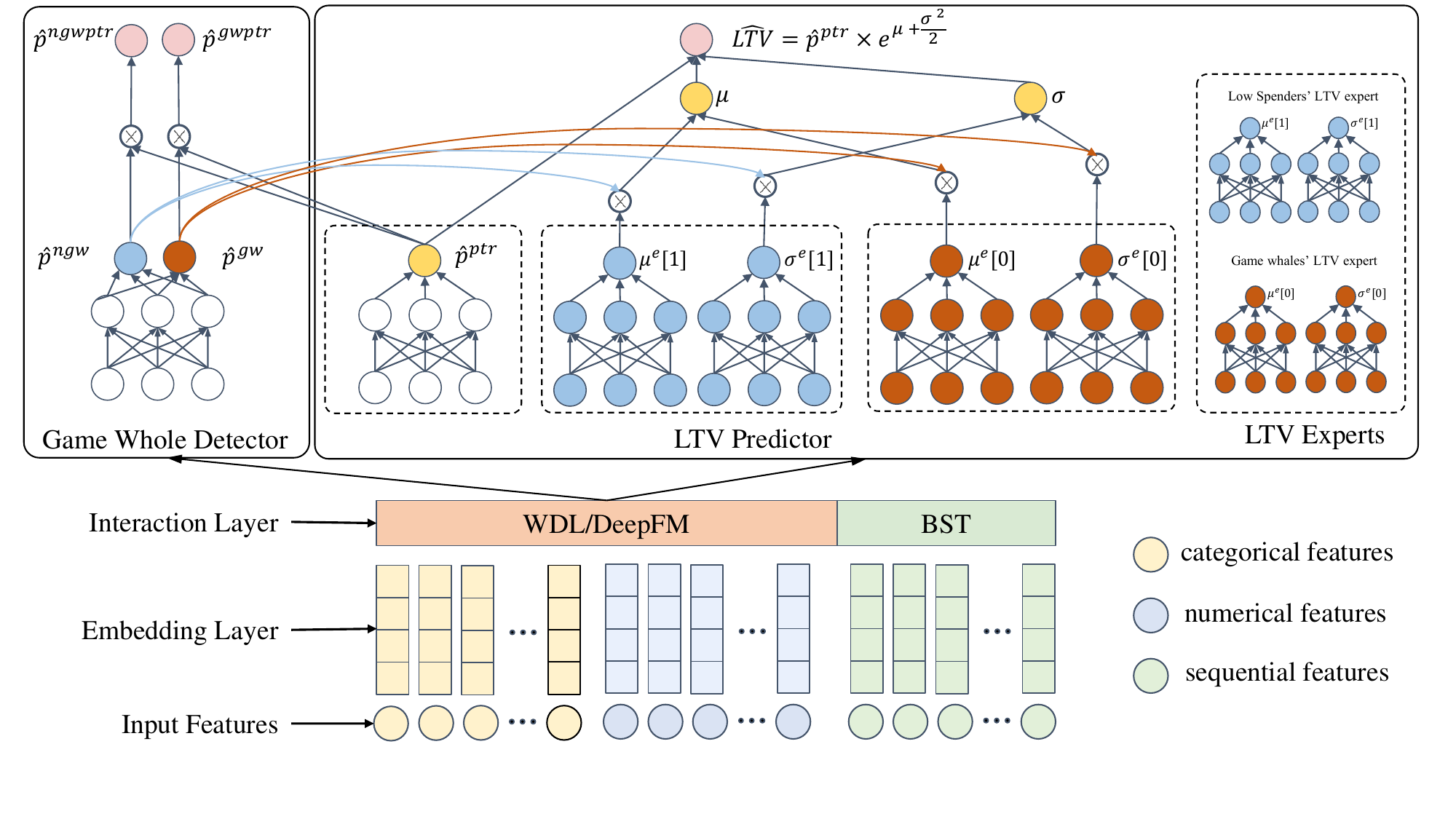}
    \vspace{-2em}
    \caption{The overview of ExpLTV.}
    \vspace{-1em}
    \label{fig:frame}
\end{figure*}
\subsection{Overview of ExpLTV}
Our proposed ExpLTV consists of two key components to perform game whale detection and LTV prediction respectively. In the game whale detection task, we carefully design two auxiliary tasks with sufficient supervisory signals to eliminate the impact of SSB and DS problems. To achieve satisfactory performance of LTV prediction, we adopt~\cite{wang2019deep} as the main building block of each LTV expert model, in which the distribution of LTV is modelled as the zero-inflated lognormal (ZILN) distribution, since it is capable of handling the extreme large LTV labels i.e., game whales' LTV values. Note that these two components are mutually enhanced with each other. Specifically, the probability of a user being classified as game whale and low spender via GW detector is taken as a weight to route each user into the right LTV expert, while the purchase rate estimator learned in both two tasks can learn the task relationships, and thus boost the performance of GW detector. In what follows, we will introduce each component in details.   
\subsection{Embedding Layer}
As illustrated in Figure \ref{fig:frame}, we first utilize a fully connected embedding layer to convert the feature vector $\mathbf{x}_u$ into low-dimensional dense representation denoted as lower-level embeddings:
\begin{equation}
    \mathbf{e}_u = \mathbf{M} \mathbf{x}_u, \forall u\in \mathcal{U},
\end{equation}
where $\mathbf{M} \in \mathbb{R}^{m \times d}$ is the feature transformation matrix and $d$ is the dimension of $\mathbf{e}_u$. In each forward iteration, to obtain the upper-level embeddings denoted as $\mathbf{e}_u^*$, the interaction layer $IntLayer(\cdot)$ is designed to preserve the feature interaction information from the lower-level embeddings $\mathbf{e}_u$. Notably, the embedding of sequence features is transformed via sequential $IntLayer(\cdot)$ such as BST~\cite{chen2019behavior}, while non-sequence features will be learned via general encoding methods (i.e., deepFM~\cite{guo2017deepfm} or WDL~\cite{cheng2016wide}). The upper-level embeddings can be represented as:
\begin{equation}
    \mathbf{e}_u^* = IntLayer(\mathbf{e}_u), \forall u\in \mathcal{U},
\end{equation}
where $d_1$ is the dimension of $\mathbf{e}_u^*$.
\subsection{Game Whale Detection}

The most straightforward way to detect game whale is to train a supervised classifier that can find the meaningful mapping between the user's upper-level embedding $\mathbf{e}^*_u$ and identity label $g_u$. However, such solution that simply divides the users into two subgroups (i.e., general users and game whales) based on the threshold value, is infeasible to correctly reflect the order of users by their monetary value. Though the classifier is well-trained, it is not only unable to distinguish non-computation users and low spenders, but also unable to assist the advertising platforms to spot the most valuable users under the limited advertising budget. Therefore, we consider the GW detection as a regression task trained by ground truth $p_u^{gwptr}$ (defined in Section \ref{df:definition}). Based on the data analysis of our real-world logs, only about $0.7\%$ of users are labeled as game whales, which inherently causes data sparsity problem. Intuitively, we found that the data volume of purchase behavior (i.e., about $11\%$ of total users) that is pre-required behavior of being a game whale is much larger. As such, a practical solution to eliminate the data sparsity problem in GW detection is to bind the purchase behavior modelling as an auxiliary task into the GW detection task. Specifically, given the training dataset $\mathcal{D} = \{(\mathbf{x}_u, LTV_u, p^{gwptr}_u): u \in \mathcal{U}\}$, we first model the purchase probability estimator $f_{ptr}(\cdot)$ as a deep neural network (DNN) that inputs a learned upper-level embedding $\mathbf{e}^*$, and computes predicted purchase probability $\hat{p}^{ptr}$, which can be represented as:
\begin{equation}
    \hat{p}_u^{ptr} = p(s_u = 1|\mathbf{x}_u),
\end{equation}
Then, user $u$'s game whale and purchase probability $p_u^{gwptr}$ can be computed following the Bayes' theorem:
\begin{equation}\label{eq:conditional}
\begin{split}
    \hat{p}_u^{gwptr} &= p(g_u=1, s_u=1|\mathbf{x}_u)\\
    &= p(s_u = 1|\mathbf{x}_u) \times p(g_u = 1|s_u = 1, \mathbf{x}_u)\\
    &= \hat{p}_u^{ptr}\times\hat{p}_u^{gw},
\end{split}  
\end{equation}
Note that Eq (\ref{eq:conditional}) holds due to the fact that purchase behavior must be occurred for the game whales. 

Furthermore, as users' LTV values often demonstrate skewed distributions with a long tail, low spenders are always account for the majority of all spenders, which illustrates the importance of low spenders in LTV prediction. Thus, a critical innovation in our approach is that our proposed GW detector is able to refine the low spenders from the non-consumption users. Specifically, we propose a novel DNN-based GW detector $f_{gwd}(\cdot)$ that computes a 2-dimensional probability distribution vector $\hat{\mathbf{y}}$ via the final softmax layer. In $\hat{\mathbf{y}}$, we let the first element $\hat{\mathbf{y}[0]}$ represent the conditional probability $\hat{p}_u^{gw}$, then $\hat{\mathbf{y}[1]}$ denoted as $\hat{p}_u^{ngw}$ is the conditional probability of being identified as a low spender, given that $u$'s purchase behavior is observed. Based on Bayes' theorem, the user $u$'s probability of being general users (i.e., low spenders or non-consumption users) $\hat{p}_u^{ngwptr}$ can be formulated as: 
\begin{equation}
    \begin{split}
        \hat{p}_u^{ngwptr} \\
        =& p(g_u=1, s_u=0|\mathbf{x}_u)+p(g_u=0, s_u=1|\mathbf{x}_u)\\
        &+p(g_u=0, s_u=0|\mathbf{x}_u)\\
    =&p(s_u = 0|\mathbf{x}_u) + p(s_u = 1 | \mathbf{x}_u)p(g_u=0|s_u=1, \mathbf{x}_u)\\
    =& (1 - \hat{p}_u^{ptr}) + \hat{p}_u^{ptr} \times \hat{p}_u^{ngw}, 
    \end{split}
\end{equation}
where $\hat{p}_u^{ngw}$ can be treated as the main indicator for identifying low spenders. 

With the aforementioned operations, we can obtain that $\hat{p}_u^{gw}$ ($\hat{p}_u^{ngw}$) is the intermediate variable of computing $\hat{p}_u^{gwptr}$ ($\hat{p}_u^{ngwptr}$) that is derived over the entire input space $\mathbf{X}$. To avoid the sample selection bias~\cite{ma2018entire} problem, an intuitive way is to simultaneously model related factor $\hat{p}_u^{gwptr}$ $(\hat{p}_u^{ngwptr})$ instead of $\hat{p}_u^{gw}$ $(\hat{p}_u^{ngw})$ by employing a multi-task learning framework. To achieve this, the GW detection loss is formulated as: 
\begin{equation}\label{eq:gwd}
    \mathcal{L}_{GWD} = \sum_{u \in \mathcal{D}}l_1(s_u, f_{ptr}(\mathbf{x}_u;\Theta_{ptr})) + D_{KL}(\mathbf{y} || \hat{\mathbf{y}}),
\end{equation}
where $\Theta_{ptr}$ is the parameter set of estimator $f_{ptr}(\cdot)$, $l_1(\cdot)$ is cross-entropy loss function and $\mathbf{y}$ is the concatenation of $p_u^{gwptr}$ and $p_u^{ngwptr}$. Note that $p_u^{ngwptr} = 1 - p_u^{gwptr}$ based on the definition. $D_{KL}$ is KL-Divergence loss function that works as a strict constraint to narrow down the distance of ground truth distributions and resulted distributions generated by the GW detector. 

\subsection{LTV Prediction}
Predicting user-level LTV is challenging but undeniably necessary for most user-centered platforms. Especially in online advertising, an LTV forecasting system built upon solid methodology plays a pivotal role in generating reasonable bidding price for each ad space on streaming media. However, users' LTV values often demonstrate long-tail distributions. Trained by those heavily imbalanced data, a conventional regressor optimized by mse loss cannot avoid to be biased towards the majority labels, while the more important minority labels (i.e., game whales' LTV values) will be underperformed. Motivated by~\cite{wang2019deep}, we transform the regression task into the task of predicting three elements, namely purchase probability $p$, mean parameter $\mu$ and standard deviation parameter $\sigma$. Each element estimator $f_{LTV_i}(\cdot)$ is designed with the same deep neural network structure that inputs a learned hidden embedding $\mathbf{e}_u^*$. 
\begin{equation}
    e^{\rhd}_i = f_{LTV_{e^{\rhd}_i}}(\mathbf{e}_u*)\text{,  } e^{\rhd}_i \in \{ p, \mu, \sigma\}, 
\end{equation}
where the activation logits units of the last layer of the DNN are sigmoid ($p$), identity ($\mu$) and softplus ($\sigma$) respectively. Moreover, most of mobile games yield significant revenue from a small percentage of game whales, which exhibit different distribution from general users. The training data with significantly large range can impede a model's capability of learning an accurate mapping from users' feature vectors to LTV values inevitably. These limitations motivate us to propose a novel LTV prediction system that can automatically allocate the right LTV expert to model users' monetary values based on users' type (i.e., high or low spenders). Since the estimator $f_{LTV_{\mu}}(\cdot)$ and $f_{LTV_{\sigma}}(\cdot)$ describe the LTV distribution, each LTV expert that contains those two components is designed by the same model structure.

Then, to automatically learn the optimal LTV experts assembling, we innovatively take $\hat{\mathbf{y}}_u$, i.e., the indicative probability of being high and low spenders as weight to compute the final distribution parameters (i.e., $\mu$ and $\sigma$):
\begin{equation}
\label{eq:agg}
\begin{split}
     \mu &= \hat{\mathbf{y}}_u\cdot\mathbf{\mu}^{e} = \hat{p}_u^{gw}\mu^{e}[0]+\hat{p}_u^{ngw}\mu^{e}[1]\\
     \sigma &= \hat{\mathbf{y}}_u\cdot\mathbf{\sigma}^{e} = \hat{p}_u^{gw}\sigma^{e}[0]+\hat{p}_u^{ngw}\sigma^{e}[1],
\end{split}
\end{equation}
where $\mu^e$ and $\sigma^e$ is the aggregation of the expert outputs, namely $\mu^e = \mu^1 \oplus \mu^2$ and $\sigma^e = \sigma^1 \oplus \sigma^2$. 
With the aforementioned operations, we can correspondingly compute $\widehat{LTV}_u$ as:
\begin{equation}
     \widehat{LTV}_u = p \cdot e^{\mu+\frac{{\sigma}^2}{2}},    
\end{equation}

Finally, we adopt the zero-inflated lognormal (Ziln) loss~\cite{wang2019deep} that is designed to handle the zero and extremely large LTV labels to optimize our LTV prediction task:
\begin{equation}
    \mathcal{L}_{LTV} = l_1(\mathbb{I}_{s_u>0}; p) + [ \mathbb{I}_{s_u>0}\log(s_u\sigma\sqrt{2\pi})+\frac{(\log s_u - \mu)^2}{2\sigma^2}], 
\end{equation}
where the first term is the cross entropy loss used to optimize estimator $f_{LTV_p}(\cdot)$, and the second term ( known as $\mathcal{L}_{lognormal}$ ) is a regression loss to quantify the prediction loss.


\subsection{Model Training}
In this section, we define the loss function of ExpLTV for model training. It is worth mentioning that estimator $f_{LTV_{p}}$ and $f_{ptr}$ designed for two tasks share the same model parameters in our model. The reason is that shared-model design can not only reduce the computation resource, but also boost performance for both GW detection and LTV prediction tasks by capturing the mutual knowledge from different perspectives. As such, we remove the first cross-entropy loss from Eq (\ref{eq:gwd}). As all components of our model are end-to-end differentiable, we combine their losses and use joint learning to optimize the following objective function:
\begin{equation}\label{eq:loss}
    \mathcal{L} = \mathcal{L}_{GWD} + \lambda \mathcal{L}_{LTV}
\end{equation}
\section{Experiments}
In this section, we first outline the evaluation protocols for our model and then conduct experiments on three industry datasets to evaluate the performance of our model. Particularly, we aim to answer the following research questions (RQs) via experiments:
\begin{itemize}
    \item \textbf{RQ1}: Is our model the new state-of-the-art in the LTV prediction task?
    \item \textbf{RQ2}: How does our model perform when detecting game whales compared with baseline methods?
    \item \textbf{RQ3}: Can we verify our contribution via the visualization method?
    \item \textbf{RQ4}: How does our model benefit from each key component?
    \item \textbf{RQ5}: How do the hyper-parameters affect the performance of our model in different tasks?
    
\end{itemize}
\subsection{Experimental Datasets}
To validate the performance of our proposed model in two tasks, we conducted experiments on three industrial datasets that are collected from Tencent Mobile Games. The user attributes in each dataset contain numerical features (e.g., age), categorical features (e.g., gender) and sequential behaviours features (e.g., purchase records). Then, we process each dataset by taking users' $T$-day cumulative consumption records as labelled LTV values (i.e., $LTV_u$). Note that an accurate model requires access to a large number of users with $LTV_u$ available, i.e., model starts using online service at least $T$ days ago. Due to the frequent changes in app updates, market and user base, a time-sensitive (i.e., a smaller $T$ day LTV) strategy is the best practice in game advertising. Thus, we set $T = 7$ in our case. As the gathered datasets are time-dependent, we take the first 40-day dataset to train the model, 3-day dataset to validate and 3-day dataset to evaluate the performance. The main statistics of our datasets are shown in Table\ref{tab: stat}.
\begin{table}[t]
\centering
\caption{Dataset Statistics.}
  \scalebox{0.99}{%
  \begin{tabular}{|c|c|c|c|}
    \hline
     Dataset&\#Users&\%Game Whales&Purchase Rate\\
     \hline
     \hline
     GAME A&292864&0.459\%&11.8\%\\
     GAME B&135168&0.675\%&7.5\%\\
     GAME C&221184&1.21\%&14.4\%\\
     \hline
  \end{tabular}}
  \label{tab: stat}
  \vspace{-2em}
\end{table}
\subsection{Evaluation Protocols}
To evaluate the effectiveness of our LTV prediction model, we adopt two popular metrics, i.e., AUC and normalized GINI (GINI). Larger values indicate better accuracy. AUC measures the performance of estimator $f_{ptr}(\cdot)$ that is designed to classify the consumption users and non-consumption users. Similarly, the GINI is purely based on the ranks of the predictions. In our case, the predicted LTV values are used as an essential factor in ad bidding, thus we follow~\cite{wang2019deep} to quantify the ranking accuracy of our model in terms of GINI. 

For GW detection, we leverage widely-used ranking metric Recall@K (R@K). Suppose we select top-K users as the most possible game whales based on the predicted probability of being identified as game whale (i.e., $\hat{p}^{gw}$), R@K is the fraction of selected game whales (i.e., $\{u |u \in GWD(\mathcal{D}, \Theta_1), g_u = 1\}$) out of all the game whales (i.e., $\{u |u \in \mathcal{U}, g_u = 1\}$). Correspondingly, larger R@K represents stronger GW detection effectiveness. 
\subsection{Baselines}
We compare our model with the following baselines on two tasks, where only the first two are trained with MSE loss and others are trained with Ziln loss. Notably, for those baselines that are only designed for LTV prediction task, top-K possible game whales are selected based on the predicted LTV values (i.e., $\widehat{LTV}$) to evaluate the performance in GW detection. Moreover, since TSUR~\cite{xing2021learning} and Marfnet~\cite{yang2023feature} are specifically designed to have a better study of user representation that is applicable in our framework, they are not selected as comparable methods in this paper. 
\begin{itemize}
    \item \textbf{Kuaishou} (KS)~\cite{li2022billion}: This work aims to deal with the complex and imbalanced distribution of LTV values by proposing a novel MDME model. 
    \item \textbf{WhalesDetector} (WD)~\cite{chen2018customer}: It uses a three-layer CNN to predict the LTV values, and then detect the valuable users based on the results.
    \item \textbf{WDL}~\cite{cheng2016wide}: It is proposed to model both low- and high-order feature interactions.
    \item \textbf{DeepFM}~\cite{guo2017deepfm}: The deep FM combines the FM~\cite{rendle2010factorization} and the deep neural network to model pair-wise feature interactions.
    \item \textbf{DCN}~\cite{wang2017deep}: A novel cross network is proposed to explicitly model feature interaction.
    \item \textbf{Ziln Loss} (ZL)~\cite{wang2019deep}: In this work, a novel zero-inflated lognormal loss is designed to handle the imbalanced regression problem.
    \item \textbf{DNN-based regressor} (SimGW): It is an upgrade version of ZL by adding a DNN-based regressor to detect GW users. 
    \item \textbf{DNN-based classifier} (SimGW2): It is an upgrade version of ZL by combining a DNN-based binary classifier to detect GW users.
\end{itemize}

\subsection{Parameters Settings}
In our model, we set the latent dimension $d$, learning rate and batch size to $8$, $0.0001$ and $128$ respectively. Model parameters are randomly initialized using Gaussian distribution. All estimator $f_{x}(\cdot)$ is formulated as a 2-layer deep neural network with 8 hidden dimensions for the hidden layer. For the coefficients in loss function $\mathcal{L}$, we set $\lambda = 15$ on GAME A, and $\lambda = 10$ on both GAME B and GAME C. 
\subsection{LTV Prediction Effectiveness (RQ1)}
\begin{figure}[t!]
\centering
\begin{tabular}{ccc}
     \multicolumn{3}{c}{\includegraphics[scale=0.375]{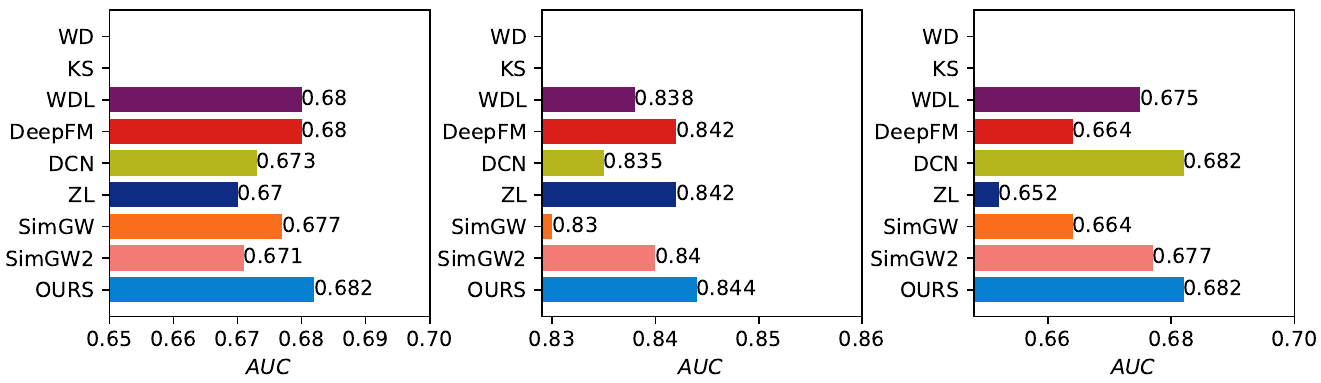}}\\
     \multicolumn{3}{c}{(a) AUC results on GAME A, GAME B and GAME C.}\\
     \multicolumn{3}{c}{\includegraphics[scale=0.335]{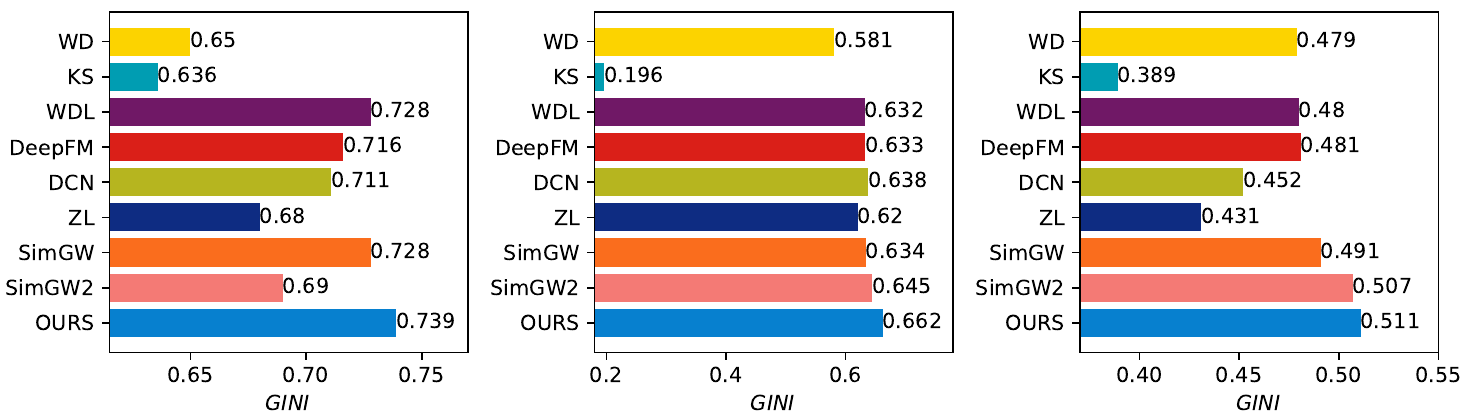}}\\ 
     \multicolumn{3}{c}{(a) GINI results on GAME A, GAME B and GAME C.}\\
\end{tabular}
\vspace{-1em}
\caption{LTV prediction Performance on three datasets.}
\vspace{-2em}
\label{fig:ltv_results}
\end{figure}

\begin{table*}
\centering
\caption{Performance of GW detection on three datasets.}
  \scalebox{0.95}{%
  \begin{tabular}{|c||c|c|c|c||c|c|c|c||c|c|c|c|}
    \hline
     Dataset&\multicolumn{4}{c||}{GAME A}&\multicolumn{4}{c||}{GAME B}&\multicolumn{4}{c|}{GAME C}\\
    \hline  Method&R@500&R@1000&R@2000&R@5000&R@500&R@1000&R@2000&R@5000&R@500&R@1000&R@2000&R@5000\\
    \hline 
    \hline
   WD&0.207&0.356&0.540&0.759&0.283&0.434&0.698&0.925&0.155&0.267&0.422&0.694\\
   KS&0.069&0.264&0.448&0.701&0.151&0.226&0.321&0.679&0.087&0.175&0.291&0.607\\
   WDL&0.253&0.390&0.598&0.759&0.264&0.472&0.717&0.981&0.126&0.243&0.417&0.704\\
   DeepFM&0.276&0.437&0.575&0.782&0.302&0.528&0.698&0.981&0.117&0.257&0.451&0.699\\
   DCN&0.299&0.460&0.610&0.793&0.208&0.453&0.736&0.962&0.160&0.301&0.413&0.699\\
   ZL&0.253&0.368&0.552&0.770&0.283&0.453&0.736&0.943&0.136&0.272&0.422&0.650\\
   
   SimGW&0.287&0.425&0.586&0.782&0.283&0.434&0.698&0.943&0.145&0.229&0.372&0.670\\

   SimGW2&0.126&0.264&0.483&0.724&0.245&0.415&0.679&0.981&0.092&0.194&0.359&0.665\\
   OURS&\textbf{0.333}&\textbf{0.471}&\textbf{0.632}&\textbf{0.793}&\textbf{0.340}&\textbf{0.547}&\textbf{0.736}&\textbf{0.981}&\textbf{0.175}&\textbf{0.301}&\textbf{0.524}&\textbf{0.713}\\
   \hline
  \end{tabular}}
  \label{tab:fd}
  \vspace{0.5em}
\end{table*}
Customer Lifetime value prediction is an essential part in the success of ads bidding platforms, since the advertisers can use the predicted results to intelligently adjust the bidding price for each ads space. We summarize all models' performance on LTV prediction w.r.t. AUC and GINI with Figure \ref{fig:ltv_results}. Note that WD and KS methods are optimized by MSE loss reported by original paper and thus the AUC results are not available. Based on the experimental results, we discuss our key findings below.

Obviously, our proposed model constantly outperforms all baselines in terms of GINI by a large margin in all three datasets, demonstrating that our model is successful at promising the advertisers to offer the reasonable bidding price for each ad space with the limited marketing budget. Specifically, compared with the best baseline, our model has brought $1.5\%$, $3.8\%$ and $6.2\%$ relative improvements on GAME A, GAME B and GAME C respectively. Additionally, the compared LTV prediction models exhibit significant performance disparity in terms of the GINI. Models optimized by Ziln loss generally perform better than models optimized by MSE loss on all datasets, indicating the superiority of Ziln loss since it is able to handle the heavy-tailedness nature of LTV values and is insensitive to the extremely large values. Though KS model is proved to deal with the complex and imbalanced distribution of $T$-day LTV values (i.e., DAU) in KUAISHOU when the range of training data is limited to a small value $T$, it is not applicable in our scenario. One possible reason is that a small number of distribution experts can KS' expressiveness in modelling the extreme behaviours of game whales, while a large number may cause overfitting problem and excessive computation resources. Finally, our model, which is the most powerful for user ranking, still achieves competitive AUC results, compared with pure LTV prediction methods. It is proved that the specific design of shared model structure can enhance the performance of estimator $f_{ptr}(\cdot)$, since it can learn the mutual knowledge from two inner-related tasks.    

   
   
\subsection{Whale Users Detection (RQ2)}
\begin{figure}
    \centering
    \begin{tabular}{cc}
     \multicolumn{2}{c}{\includegraphics[scale=0.33]{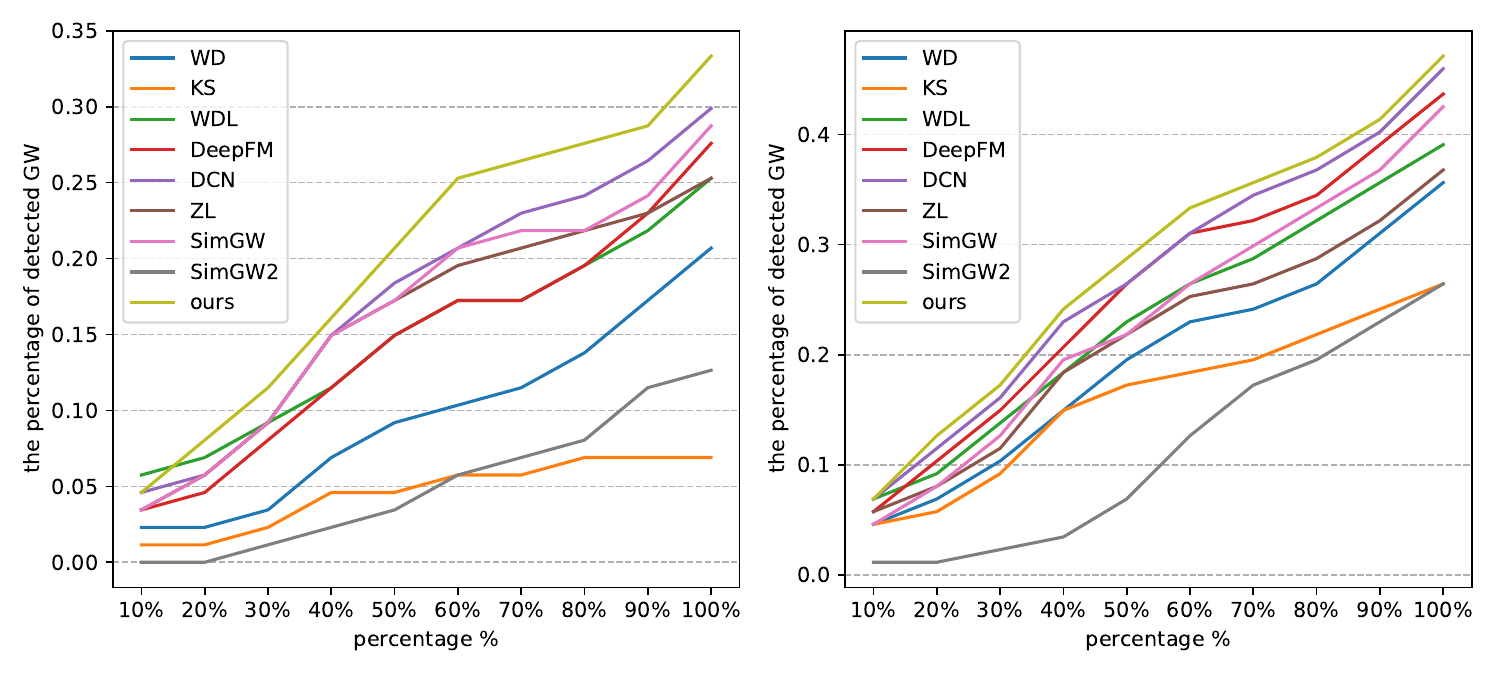}}\\
     \hspace{2.5em}(a) Search space $K = 500$ &(b)Search space $K = 1000$ \\
    \end{tabular}
    \caption{GW detection results on three datasets. Point $(x, y)$ means that $y$ GWs ranking above level $x$ are detected.}
    \label{fig:fgdw}
    \vspace{-1em}
\end{figure}

Game whale is a tiny group of users that brings the most revenue for mobile games. Hence, it is imperative for advertisers to spot this type of users among massive new users for game publishers, especially for top-grossing games. To quantitatively evaluate the effectiveness of ExpLTV in GW detection, we provide comprehensive analysis from different perspectives.  

We first report the overall detection performance of all tested methods in Table \ref{tab:fd}. Note that by increasing the searching space (i.e., the value of K), it becomes easier for detectors to retrieve game whales. The first observation we can draw is that our model is successful at detecting the game whales. In particular, when $K = 5000$, the GW detector becomes highly confident in its detected results and over $70\%$ of game whales can be spotted accurately in all datasets. Second, our model outperforms all baseline methods consistently. The improvements of ExpLTV significantly increases, with the reduced search space. On GAME A, the $11.4\%$ relative improvement of our model with $K = 500$ demonstrates that ExpLTV can capture the most valuable game whales even under the extremely limited searching budget. Furthermore, the models that achieve better results in the LTV prediction task may not still perform well in GW detection. For example, WLD and DeepFM outperform DCN in terms of GINI, while the R@K values are slightly lower. Without the awareness of game whales in the design of pure LTV models, the well-trained models are limited to perform well only on the majority of labels (i.e., low spenders' LTV values), and thus they are unable to always rank game whales at the top of the non-game whales. To prove it, we further calculate the GINI for all spenders ($GINI_{1}$) and high spenders ($GINI_{R}$). Compared with the best baseline (i.e., DCN on GAME A and GAME C,  and DeepFM on GAME B), the improvements achieve $0.074$ ($GINI_{1}$) and $0.303$ ($GINI_{R}$) on GAME A, $0.04$ ($GINI_{1}$) and $0.082$ ($GINI_{R}$) on GAME B, and $0.054$ ($GINI_{1}$) and $0.122$ ($GINI_{R}$) on GAME C. The results further validate that the pure LTV models are impractical in the GW detection task. Moreover, SimGW that makes use of a regressor to detector game whales outperforms SimGW2 that designs the detector as a simple binary classifier, which verifies the effectiveness of our core idea.     
\begin{figure}
    \centering
    \begin{tabular}{cc}
     \multicolumn{2}{c}{\includegraphics[scale=0.37]{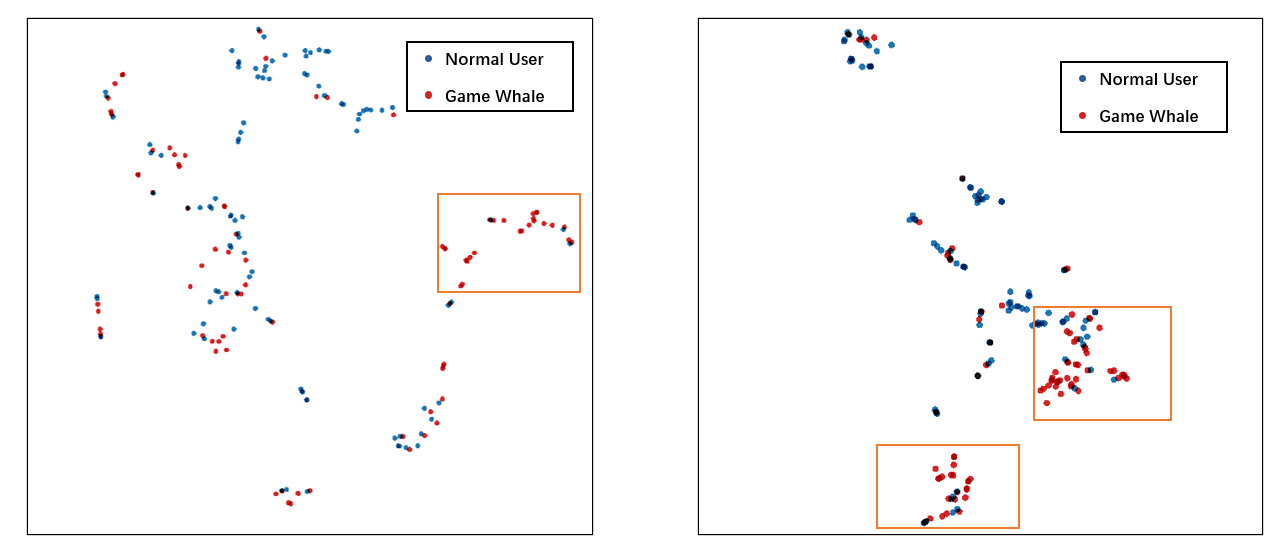}}\vspace{0.5em}\\
     \hspace{1.2em}(a) Pure LTV model i.e., ZL&\hspace{0.4em}(b) ExpLTV\\
    \end{tabular}
    \caption{Visualization of latent embeddings before and after removing game whale detection component. The visualization corresponds to the GAME A dataset. For better clarity, the number of general users is downsampled to 100.}
    \label{fig:visualize}
    \vspace{-1em}
\end{figure}

Since the main use of GW detector is to spot the most valuable users with the limited search budget, we further test the detected results' quality of all baselines from a fine-grained view. Specifically, we partition labelled game whales into ten levels based on the LTV values, and then calculate the number of detected game whales ranking above each level. Figure \ref{fig:fgdw} reports the results with $K = 500$ and $K = 1000$. Clearly, ExpLTV constantly achieves the best performance, verifying that our model can not only detect more game whales, but also can accurately catch those most valuable game whales who have more significantly impact on games' revenue.

\subsection{Visualization Results (RQ3)}
We are the first to bind GW detection as an auxiliary task to boost the LTV prediction performance especially for game whales, since the extremely large and imbalanced labels of them exhibit dramatically varied distribution from general spenders. To verify the necessity of the GW detection in ExpLTV, we visualize the upper latent embeddings $\mathbf{e}_u^*$ in ExpLTV and pure LTV prediction model i.e., ZL via t-SNE in Figure \ref{fig:visualize}. As can be told from Figure \ref{fig:visualize}, the latent embeddings forms several distinct clusters based on the type of users in ExpLTV. Specifically, two large red game whale clusters, one small and one large blue general user clusters. Though several small clusters can be observed, most of the latent embeddings are mixed together in ZL. Furthermore, only several game whales are forced to fit into the wrong clusters in ExpLTV. Hence, the discriminative embeddings encoded by useful information in our model can prove that the application of GW detection has a significant benefit in the LTV prediction.   

\subsection{Ablation Study (RQ4)}
\begin{table}
    \caption{Ablation test results.}
    \centering
    \scalebox{0.89}{%
    \begin{tabular}{|p{1.1cm}<{\centering}|p{2cm}<{\centering} |p{1cm}<{\centering}|p{1cm}<{\centering}||p{1cm}<{\centering}|p{1cm}<{\centering}|}
        \hline
         \multirow{2}{*}{Dataset}&\multirow{2}{*}{Variant}&\multicolumn{2}{c||}{Ltv Prediction Task}&\multicolumn{2}{c|}{GW Detection task}\\
          \cline{3-6}
         &&AUC&GINi&R@500&R@1000\\
         \hline
          \multirow{3}{*}{GAME A}&ExpLTV-ne&0.673&0.577&0.333&0.425\\
         &ExpLTV-nssb&0.680&0.739&0.276&0.459\\
         &ExpLTV-sp&0.671&0.690&0.287&0.460\\
         &ExpLTV&\textbf{0.682}&\textbf{0.739}&\textbf{0.333}&\textbf{0.471}\\
         \hline
         \multirow{3}{*}{GAME B}&ExpLTV-ne&0.835&0.624&0.320&0.528\\
         &ExpLTV-nssb&0.842&0.658&0.302&0.434\\
         &ExpLTV-sp&0.844&0.646&0.302&0.528\\
         &ExpLTV&\textbf{0.844}&\textbf{0.662}&\textbf{0.340}&\textbf{0.547}\\
         \hline
         \multirow{3}{*}{GAME C}&ExpLTV-ne&0.677&0.488&0.170&0.291\\
         &ExpLTV-nssb&0.676&0.513&0.155&0.296\\
         &ExpLTV-sp&0.677&0.515&0.160&0.300\\
         &ExpLTV&\textbf{0.682}&\textbf{0.517}&\textbf{0.175}&\textbf{0.301}\\
         \hline
    \end{tabular}}
    \label{tab:ab}
    \vspace{-1em}
\end{table}
To better understand the performance gain from different major components proposed in our model, we implement several degraded versions of ExpLTV for ablation analysis. Table \ref{tab:ab} summarizes the outcomes in two tasks in terms of AUC, GINI and R@K. In what follows, we describe all variants and analyse the effectiveness of corresponding model components. 

\textbf{Removing LTV Experts (ExpLTV-ne)}. In our model, the predicted LTV values are generated by aggregating the outputs of each LTV expert via Eq.(\ref{eq:agg}). To testify the usefulness of LTV experts, we only retain one LTV expert and the result is treated as the final output of the LTV prediction, then we use joint learning to optimize the final objective function. As can be inferred from Table \ref{tab:ab}, compared with other methods that are designed with multiple LTV experts (i.e., ExpLTV-nssb and ExpLTVs-sp), ExpLTV-ne has the worst performance in LTV prediction, which validates the importance of making full use of LTV experts to capture the distribution differences between high and low spenders. Additionally, a slight performance drop can be observed in the GW detection task. One possible reason is that the user-specific LTV expert by design can boost the LTV prediction accuracy of game whales, which correspondingly enhances the GW detection effectiveness.
\begin{figure*}
    \centering
    \begin{tabular}{cccc}
     \multicolumn{4}{c}{\includegraphics[scale=0.68]{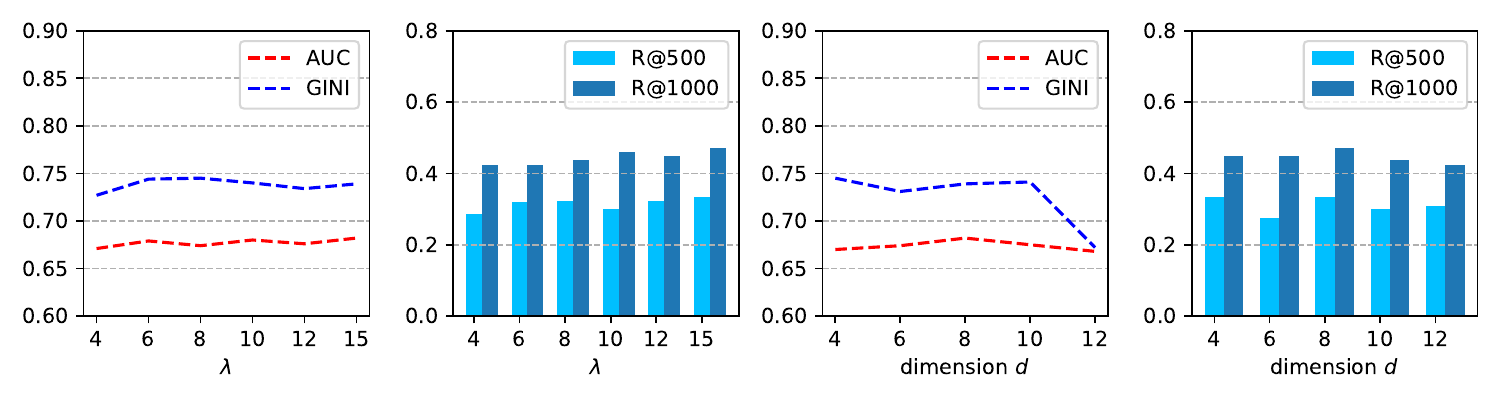}}\vspace{-0.5em}\\
     \hspace{4.5em}(a) LTV Prediction& \hspace{5.5em}(b) GW Detection& \hspace{5em}(c) LTV Prediction& \hspace{2.5em}(d) GW Detection\\
     \end{tabular}
     \vspace{-1em}
    \caption{Parameter sensitivity results w.r.t. $\lambda$ and dimension $d$ on GAME A.}
    \vspace{-1em}
    \label{fig:sens}
\end{figure*}

\textbf{Removing Sequential behaviours learning (ExpLTV-nssb)}. This variant disables the sequential behaviour "$convert\rightarrow purchase\\ \rightarrow game whales$" learning by first removing estimator $f_{ptr}(\cdot)$, and then rewriting $\hat{\mathbf{y}} = [\hat{p}^{gw}, \hat{p}^{ngw}]$, i.e., $\hat{p}_u^{gw}$ ($\hat{p}^{ngw}$) is the approximation of $\hat{p}_u^{gwptr}$ ($\hat{p}^{ngwptr}$). As ExpLTV-nssb no longer takes the purchase rate prediction as an auxiliary task into the GW detection to counteract the data sparsity and sample selection bias problem, it suffers from inferior performance in GW detection. For R@500, the significant performance drop reaches $20.7\%$ on GAME A, $12.6\%$ on GAME B and $12.9\%$ on GAME C respectively. In addition, the slight performance drop of GINI validates that the shared purchase rate estimator performing for two tasks by design can contribute to the expressiveness of LTV experts. Thus, the novel Sequential behaviours learning showcases its strong contribution to the performance gain in our model.    


\textbf{Using Individual Purchase Probability Estimator for Each Task (ExpLTV-sp)}. A crucial difference between ExpLTV and ExpLTV-sp is the design of the purchase probability estimator $f_{ptr}(\cdot)$. Since the LTV prediction and GW detection are two closely related tasks, the shared estimator by design can capture the inner task relationships. As a core part in ExpLTV, we further testify the efficacy of shared estimator $f_{ptr}(\cdot)$. ExpLTV constantly outperforms ExpLTV-sp in both two tasks across all datasets. On GAME A, the performance decrease reaches $1.64\%$ in AUC, $7.1\%$ in GINI, $16\%$ in R@500 and $2.4\%$ in R@1000. The results further validate that the estimator trained from two views can learn the mutual knowledge, and thus it can boost the performance of both two tasks.  

\subsection{Hyper-Parameter Sensitivity (RQ5)}

To answer RQ4, we further investigate the performance fluctuations of ExpLTV with two varied hyperparameters on GAME A, namely trade-off $\lambda$ between LTV prediction loss and GW detection loss in Eq (\ref{eq:loss}), and latent dimension $d$. Based on the standard setting $\{d = 8, \lambda = 15\}$ of ExpLTV, we tune the value of one hyperparameter while keeping the other unchanged, and report the new results of two tasks achieved in Figure \ref{fig:sens}. Specifically, We record the performance differences by plotting AUC and GINI for LTV prediction, while demonstrating R@500 and R@1000 for GW detection. 

\textbf{Impact of $\lambda$}
We study our model's sensitivity to the value of $\lambda$ in $\{4, 6, 8, 10, 12, 15\}$ that controls the trade-off between LTV prediction and GW detection. Within our expectation, as $\lambda$ increases from $4$ to $15$, there is a slight performance drop in LTV prediction, while an upward trend can be observed in GW detection. Luckily, altering this coefficient has less impact on the LTV prediction. Thus, setting $\lambda = 15$ is sufficient for improving the accuracy of GW detection, while ensuring the satisfactory performance of LTV prediction of ExpLTV. 

\textbf{Impact of $d$}. We vary dimension $d$ in $\{4, 6, 8, 10, 12\}$. Generally, the value of dimension $d$ directly controls our models' expressiveness. As $d$ increases from $4$ to $10$, a fluctuating growth can be observed in ExpLTV performance. However, when $d$ exceeds $10$, the performance improvement tends to stop. As can be inferred from Figure \ref{fig:sens} (c) and (d), our model with $d = 8$ can achieve the best  or second best results in both LTV prediction and GW detection, and thus we set $d = 8$ to achieve a balance between the accuracy of both two tasks.     
\section{Related Work}
\textbf{LTV Prediction.} Understanding the total revenue that the business can expect from a customer is important in user acquisition~\cite{kumar2004customer,venkatesan2004customer,pfeifer2005customer,yang2023feature}. In the literature, many LTV prediction models have been proposed, which can be categorised as probability-based, machine learning- based and deep learning-based methods. The probability-based methods assume the purchase behaviors as a probability distribution, and then propose the probabilistic generative models for predicting the LTV values~\cite{gupta2006modeling,fader2005rfm,fader2005counting,schmittlein1987counting}.~\cite{fader2005rfm} proposes a stochastic model that links the RFM paradigm~\cite{mccarty2007segmentation} with LTV values. The machine learning-based methods are proposed to learn a mapping between hand-crafted features and monetary value of game players via machine learning techniques~\cite{vanderveld2016engagement, drachen2018or, chamberlain2017customer, gupta2006modeling}. For example, ~\cite{drachen2018or} adopts a random forest model to predict user-level LTV value in Groupon. Recent years have witnessed the successful development of deep learning-based techniques in LTV prediction.~\cite{chen2018customer} designs a convolutional neural network in LTV prediction to perform a better modelling of temporal representations. ZL~\cite{wang2019deep} models the distribution of LTV values as Ziln distribution to capture the long-tail nature of training data. Meanwhile, the Ziln loss can be used in deep learning-based neural networks. TSUR~\cite{xing2021learning} is designed to learn a more stable user representation by utilizing wavelet transform and Graph Attention Network, which can alleviate the volatility and sparsity problems of monetary values. In~\cite{yang2023feature}, a feature
missing-aware routing-and-fusion network (MarfNet) is proposed to reduce
the effect of the missing features while training. Recently,~\cite{li2022billion} is developed to predict users' DAU in KUAISHOU. In~\cite{li2022billion}, the LTV distribution is divided into multiple sub-distributions trained via distribution experts. Inherented the limitations of mse loss, ~\cite{li2022billion} can be proved to success when the range of monetary values is limited to a small value, and thus our scenario impedes its model expressiveness. Most of existing research efforts mainly focus on improving the performance of LTV prediction by enhancing the feature representation, while the potential of integrating GW detection and LTV prediction into a unified framework to make most of their beneficial relationship is always ignored. These limitations motivate us to propose ExpLTV that is able to achieve superior performance in both LTV prediction and GW detection. 

\textbf{Multi-Task Learning.} Multi-Task learning (MTL) is a widely used training paradigm in machine learning~\cite{zhang2021survey, sener2018multi, zhang2022pipattack,zhang2020gcn}. It aims to capture the inner relationships among multiple tasks to improve the performance of each task. The supervised MTL can be classified into five main categories: feature learning-based~\cite{caruana1998multitask,liao2005radial}, low rank-based~\cite{ando2005framework,zhang2005learning}, task clustering-based~\cite{thrun1996discovering,bakker2003task}, task relation learning-based~\cite{evgeniou2005learning,kato2007multi} and decomposition-based~\cite{jalali2010dirty,gong2012robust} approaches. In recent years, many sort of works are successful in solving sample selection bias (SSB) and data sparsity (DS) problems by utilization of MTL in conversation rate prediction (CVR). Specifically, SSB is a bias caused by the varied distribution of training space and inference space.~\cite{ma2018entire} is the first to propose an entire space multi-task model (ESMM) to eliminate SSB and DS problems in CVR task. In ESMM, instead of directly optimizing CVR task, two auxiliary tasks of predicting the post-view click-through rate (CTR) and post-view click-chrough\& conversion rate (CTCVR) are introduced. Inspired by ESMM, ESM${^2}$~\cite{wen2020entire} decomposes "$impression\rightarrow purchase$" into several intermediate behaviours as"$impression\rightarrow click \rightarrow D(O)Action\rightarrow purchase$". Based on the novel sequential behavior decomposition, CVR prediction is optimized by modelling multiple auxiliary tasks instead. Similarly, the SSB and DS problems exist in game whale detection. In our work, we form a new sequential behaviour "$convert\rightarrow purchase\rightarrow game whales$", and then trains two decomposition tasks i.e., purchase rate prediction and game whale \& purchase rate prediction via multi-task learning. Consequently, investigating by the entire samples and abundant auxiliary supervisory signals, 
ExpLTV can efficiently address the SSB and DS issues. 

\section{Conclusion}
In this paper, we propose a novel multi-task framework named ExpLTV to perform LTV prediction and game whale detection. By investigating the beneficial relationship of LTV prediction and game whale detection, these two tasks can exert mutually. In game whale detection, the carefully designed DNN-based detector is expected to precisely refine game whales and low spenders, which can be treated as a gating network to decide the optimized patterns of LTV experts assembling. Meanwhile, the purchase rate estimator trained by the LTV predictor is used in the GW detector as an auxiliary task to eliminate SSB and DS problems. The extensive experiments conducted on three industrial datasets confirm the effectiveness of ExpLTV over the state-of-the-art baselines on both LTV prediction and GW detection tasks.  
\clearpage
\bibliographystyle{ACM-Reference-Format}
\bibliography{sample-base}

\end{document}